\documentclass[preprint,psfig]{aastex}
\begin{document}

\title{Constraints on $\Omega_B$, $\Omega_m$ and $h$ from  MAXIMA and BOOMERANG }
\author{T. Padmanabhan}
\affil{{Inter-University Centre for Astronomy \& Astrophysics, \\
Post Bag 4, Pune 411007, India} }
\email{paddy@iucaa.ernet.in}
\and
\author{Shiv K. Sethi}
\affil{Mehta Research Institute,\\ Chhatnag Road, Jhusi, Allahabad
211 019, India}
\email{sethi@mri.ernet.in}

\begin{abstract}
  We  analyse the BOOMERANG and MAXIMA results in the context of
   models with $\Omega_{\rm Total} = 1$ and $n_s =
  1$.  We attempt to constrain three other parameters---$h$, $\Omega_B$, and
  $\Omega_m$---from these observations. We show that:  (a)
  the value of $\Omega_B h^2$ is too high to be compatible with
  primordial nucleosynthesis observations at 95\% confidence level (b)
  universe with age greater than
  $ 12 \, \rm Gyr$ is ruled out at 95\% confidence level and
  (c) the value of $\Omega_m h$ is too high
  to be compatible with the shape of the power spectrum of
  gravitational clustering. In effect, our analysis shows that
   models with $\Omega_{\rm Total} = 1$ and $n_s =  1$ are
 ruled out by  BOOMERANG and MAXIMA  observations.
  \end{abstract}

Precise determination of CMBR anisotropies has long been expected to
give accurate values of cosmological parameters (see e.g. Bond,
Efstathiou \& Tegmark 1997 and references therein). These
cosmological parameters include parameters of background FRW model
($\Omega_{\rm Total}$, $\Omega_\Lambda $, $h$, $\Omega_B$, etc.),
parameters that determine the formation of structure in the universe
( $\sigma_8$, scalar spectral index $n_s$, etc.),  and
the parameters related to the re-ionization of the universe (the optical depth to
the last scattering surface, $\tau$ etc.).

While the future
experiments MAP and Planck \footnote{For details  see
  {\tt http://map.gsfc.nasa.gov} and 
  {\tt http://astro.estec.esa.nl/SA-general/Projects/Planck}},
largely owing to their all-sky coverage,
are expected to determine most of these parameters with a few percent
accuracy (Jungeman {\it et al. \/} 1996, Zaldarriaga {\it et al.}
1997, Prunet {\it et al.} 1999), recent observations have already begun to give important
clues about  some of these parameters (Miller {\it et al. \/} 1999,
Mauskopf {\it et al.} 2000, Netterfiled {\it et al.} 1997, for a
summary of observation up to 1998 and parameter estimation from these
observations, see Lineweaver \& Barbosa 1998). Recent balloon experiments
BOOMERANG and MAXIMA reported CMBR anisotropy measurements at
angular scales between $\simeq 10^\circ$ and $ \simeq 10'$ (de Bernardis {\it
  et al. \/} 2000, Hanany {\it et al. \/} 2000). These
experiments observed nearly $1 \%$ of the sky with angular
 resolution  $\simeq 10'$.   For both these experiments the cosmic
 variance was small enough  (owing to the sky coverage)
 to determine precisely the position of the first
 Doppler peak ($\ell \simeq 200$) of the CMBR anisotropies. Both
 BOOMERANG and MAXIMA results gave strong evidence that 
  $\Omega_{\rm Total} = 1$ (de Bernardis {\it
    et al. \/} 2000, Hanany {\it et al. \/} 2000), which was already
  indicated by other observations (Netterfield {\it et al.} 1997).

 While the position of the first Doppler peak give unambiguous evidence about
 the geometry of the universe, determination of other cosmological
 parameters is more difficult. This is because variation in several
 different parameters give the same change in measured anisotropies,
 e.g. the height of first Doppler peak is nearly degenerate in $\Omega_B$,
 $h$,  $\Omega_\Lambda$ and $n_s$. Some of this degeneracy can be
 lifted with the measurement of anisotropies at even smaller angular
 scales.   BOOMERANG and MAXIMA probe with angular scales
 corresponding to multipoles $\ell_{\rm max} \simeq
 \{600,700\}$, respectively, which is up to or beyond the
 expected position of the second Doppler peak. Though the results of
 these experiments have not been able to find the position of the
 second Doppler peak, accurate measurement of anisotropies at such angular
 scales is expected to break some of the degeneracy which measurements
 near the first Doppler peak alone cannot.

 The BOOMERANG and MAXIMA data have been used to determine various
 cosmological parameters (Balbi {\it et al. \/} 2000, Jaffe {\it et
   al. \/} 2000, Lange {\it et al.} 2000, Tegmark \& Zaldarriaga
 2000). Combined with other independent measurements
 of cosmological parameters
 (e.g. measurement of $\Omega_B h^2$ from element abundance, 
 measurement of $h$ from nearby observations or inference about the
 values of $\Omega_\Lambda$ and $\Omega_m$ from the SN1a data, etc.)
 these data  are expected to lead to a unique picture. However, owing
 to degeneracies in parameter estimation, the value of estimated
 parameters and their errors depend sensitively on various assumption
 related to the assumed allowed range of parameters, i.e. on the
 priors on the parameters.

 In this letter, we perform a likelihood analysis on the  band-powers
 reported by the MAXIMA and BOOMERANG experiments. However, instead of
 of doing a multiple parameter analysis, we fix the values of most
 parameters from other considerations (and prejudices!) and then attempt to
 estimate just three parameters---$\Omega_B$, $h$, and
 $\Omega_m$   assuming  weak
 priors on their allowed values. In the next section, we explain our
 choice of parameters and the method we use in brief. In the third
 section, we present and summarise our results.

 \section{Cosmological parameters and CMBR data}

 Most generic models of inflationary scenario give two unique
 predictions: $\Omega_{\rm Total} = 1$ and $n_s \simeq  1$
 (see e.g. Steinhardt 1995).  The first
 of these predications is confirmed by BOOMERANG and MAXIMA. The
 analysis of COBE-DMR data is consistent with $n_s = 1$ (Bennett {\it
   et al \/} 1996, Bunn \& White 1996).  Therefore it
 is reasonable to believe that the current data is in good agreement
 with these predictions.
 We fix the value of these two parameters based on these
 considerations and use $\Omega_{\rm Total} = 1$ and $n_s \simeq 1$
 throughout.    (It should be pointed out that
 $\Omega_{\rm Total} = 1$ is a stricter requirement of inflation than
 $n_s = 1$; one gets $n_s = 1$ only for exponential inflation; see e.g. Steinhardt
 1995.) Also note that we are not concerned with the origin of the values
$\Omega_{\rm Total} = 1$ and $n_s =
1$. For example, these could arise merely from the requirement of
scale invariance
for the background universe (giving $\Omega_{\rm Total}=1$) and the 
perturbations (giving $n_s=1$) without invoking inflation --- as was
originally done by Harrison and Zeldovich, years before inflation was
invented (Harrison 1970, Zeldovich 1972).
But, of course, inflationary models made these parameter values fairly
well accepted.
Other parameters like $\Omega_\Lambda$, $h$ and
 $\Omega_B$ cannot be fixed by theoretical considerations alone.
 In our analysis we assume a non-zero $\Omega_\Lambda$ because
 recent high-z SN1a observations suggest a non-zero cosmological constant
 (Perlmutter {\it et al.} 1999, Riess {\it et al.} 1998).
 We do not consider CMBR anisotropies from tensor perturbations as
 most models of inflation give negligible contribution from tensor
 perturbations for $n_s =1$ (Steinhardt 1995).
 Re-ionization of universe can also alter
 primary CMBR anisotropies significantly. Present observations suggest
 that the universe is ionized up to $z \simeq 5$. For the CMBR
 anisotropies to be significantly altered by re-ionization, the
 minimum re-ionization redshift should lie between $10$ and $20$ (see
 e.g. Bond
 1996). Therefore, keeping the constraints from present observations in
 mind, we neglect the effect of re-ionization on the measured CMBR
 anisotropies.
 
We  fix the value of  $\Omega_m  + \Omega_\Lambda = \Omega_{\rm
   Total} = 1$, and compute the confidence levels on the best-fit
 values of $\Omega_B$ and $h$, and $\Omega_m$
 from MAXIMA and BOOMERANG observations.
 The range of parameters in which the minimum of $\chi^2$ is
 searched is: $0.01 \le \Omega_B  \le 0.15$, $0.4 \le h \le 1.1$ and
 $0.1 \le \Omega_m \le 0.95$. 

 The $\chi^2$ for the model comparison with observations is given by:
 \begin{equation}
   \chi^2 = \sum_{i = 1}^N \left ({{\cal C}_\ell^{\rm obs} - {\cal
         C}_\ell^{\rm th} \over \Delta {\cal C}_\ell^{\rm obs}}
   \right)
   \label{eq:a1}
   \end{equation}
   Here $N = 22$ (10 points from MAXIMA and 12 points from BOOMERANG),
   ${\cal C}_\ell^{\rm obs}$ are the measured band-powers and 
   ${\cal C}_\ell^{\rm th}$ are the theoretical band-powers, $\Delta
   {\cal C}_\ell^{\rm obs}$ are the errors on measured band-powers.
   We do not take into account the calibration uncertainties in our
   analysis. The
   theoretical band-powers are calculated using the CMBR Boltzmann code
   CMBFAST (Seljak \& Zaldarriaga 1996), which  gives COBE-normalized
   (normalized according to the fitting formula of Bunn \& White
   (1996)) angular power spectra. 
   Eq.~(\ref{eq:a1}) assumes that different band-powers are
   uncorrelated. This assumption is valid only if the anisotropies are
   measured over the entire sky (see e.g. Peebles 1993, Padmanabhan 1993).
   We assume lack of correlation in this work also
   because the covariance matrix of band-powers has not been made
   public yet.

  Eq.~(\ref{eq:a1}) implicitly assumes that the
 likelihood function is Gaussian in band-powers near its
 maximum. While this assumption is true in principle, in practice
 there can be significant deviation from Gaussianity near the
 maximum. Bond, Jaffe, and Knox (2000) advocate using another variable
 instead of band-powers for doing the maximum likelihood
 analysis. We do not use it here.
 However,  while quoting errors we do not use the Fisher matrix approach
 which can give meaningful results only for the Gaussian case.
 Instead  we directly give
 the confidence levels on $\Delta \chi^2$.

 \section{Results}

 Our results are shown in Figures~1 and~2.  The $\chi^2$ for 22 data
 points from BOOMERANG and MAXIMA with three  fitted parameters
 (i.e. 19 degrees of freedom) is 22.4, which is an
 excellent fit (Goodness-of-fit probability, $Q = 0.26$). The best fit values 
 and $1\sigma$ errors are:
 $\Omega_B = 0.075^{{+}0.003}_{{-} 0.015}$, $h =0.62^{{+}0.08}_{{-}
   0.02}$ and $\Omega_m=  0.86^{{+}0.04}_{{-} 0.21}$. The best fit
 model along with BOOMERANG and MAXIMA data points is shown in
 Figure~3.

 The  range of allowed $h$ is in fair agreement with the
 measurement of $h$ from local observations (for a recent summary of
 these results, see Primack 2000). Recent SN1a suggest that that the
 universe is flat with $\Omega_\Lambda + \Omega_m = 1$ with
 $\Omega_m = 0.28^{{+}0.09{+}0.05}_{{-}0.08{-}0.04}$ (1$\sigma$) .
 This is within $\Delta \chi^2 = 2$ of the value inferred by our
 analysis. 

 In Figure~1 we show the confidence levels in the
 $\Omega_B\hbox{--}h$ plane. The region bounded by the contours
 correspond to $\Delta \chi^2 \le 2.3$ and $\Delta \chi^2  \le 6.17$, which, for Gaussian errors,
 corresponds to 68~\%  (1$\sigma$) and 95.4~\% (2$\sigma$) for
 two-parameter fits. We also show the  95\%  region allowed by
 primordial nucleosynthesis observations  (for a recent review see
 Tytler {\it et al} 2000). As seen in
 the figure, the region allowed by CMBR observations is at variance
 with the predictions of primordial nucleosynthesis at at least  95\% level. This
 result is in agreement with other analyses on the same data set
 (Tegmark \& Zaldarriaga 2000, Jaffe {\it et al} 2000).

 The region corresponding  to one and two $\sigma$  in the
 $\Omega_m\hbox{--}h$ plane is shown in Figure~2.  Our results
 show that the current CMBR
 observations  favour a universe with age $\le 11 \rm Gyr$ and
 are incompatible with a universe of age $\ge 12 \, \rm Gyr$ at
 $\simeq 95 \%$ level.  Though this is on the lower side of the
 expected age of the universe from estimated ages of globular clusters,
 etc., this is not in disagreement with those observations (for a
 recent status report see Primack 2000). 
 Another important constraint on the values of $\Omega_m$ and $h$ comes
 from the shape of the power spectrum of galaxy clustering (see
 e.g. Bond 1996). These observations, within the context of a flat
 model with cosmological constant, give $\Omega_m h \simeq 0.25 \pm
 0.05$.  We plot this region in the  $\Omega_m\hbox{--}h$ plane in
 Figure~2.   If we require that $h \le 0.75$, as most recent
 measurements of $h$ suggest (Primack 2000), then CMBR observations
 exclude the region required to satisfy the galaxy clustering
 observations by more than $95\%$ confidence level.

 In  conclusion, recent BOOMERANG and MAXIMA observations, within the
 context of simplest inflationary models ($\Omega_{\rm Total} = 1$ and
 $n_s \simeq 1$),
 imply:  (1) too large a of value
 of $\Omega_Bh^2$ to be compatible with primordial nucleosynthesis
 observations, (2) an age of universe $t_0 \le 12 \, \rm Gyr$, and (3)
 too large a value of $\Omega_m h$ to be in agreement with the shape of
 the power spectrum of galaxy clustering.  In view of this, it is
 safe to conclude  that
 the  models with $\Omega_{\rm Total} = 1$ and $n_s =  1$ are
 ruled out by these  observations.

 \newpage
\begin{figure}
\figurenum{1}
\includegraphics[angle=270, width=\textwidth]{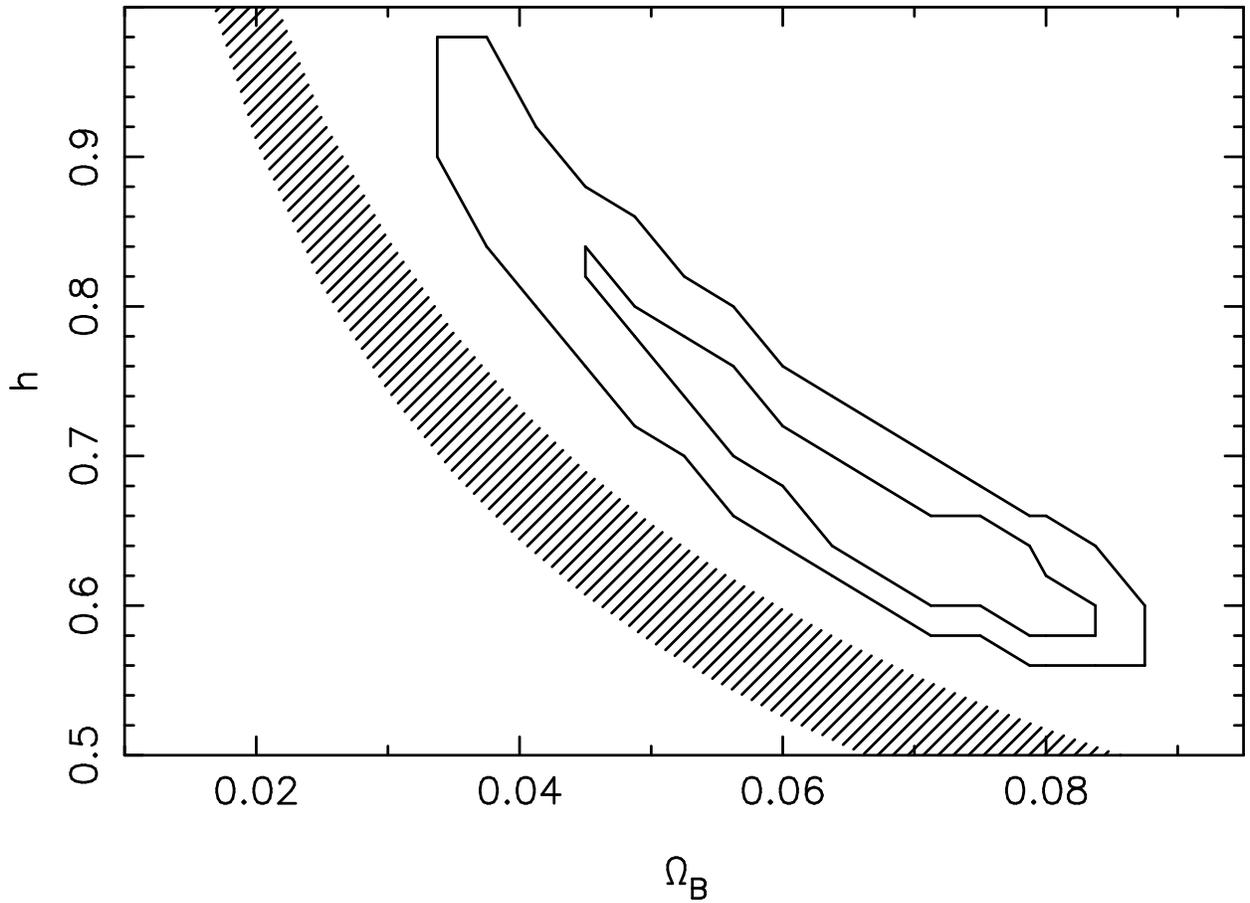}  
\caption{The contours correspond to allowed 1 and 2 $\sigma$
  regions by CMBR observations (see text for detail). The cross-hatched region
  correspond to the 95\%  region ($\simeq 2\sigma$)
  from primordial nucleosynthesis (Tytler {\it et al.} 2000). \label{fig:1}}
\end{figure}

\begin{figure}
\figurenum{2}
\includegraphics[angle=270, width=\textwidth]{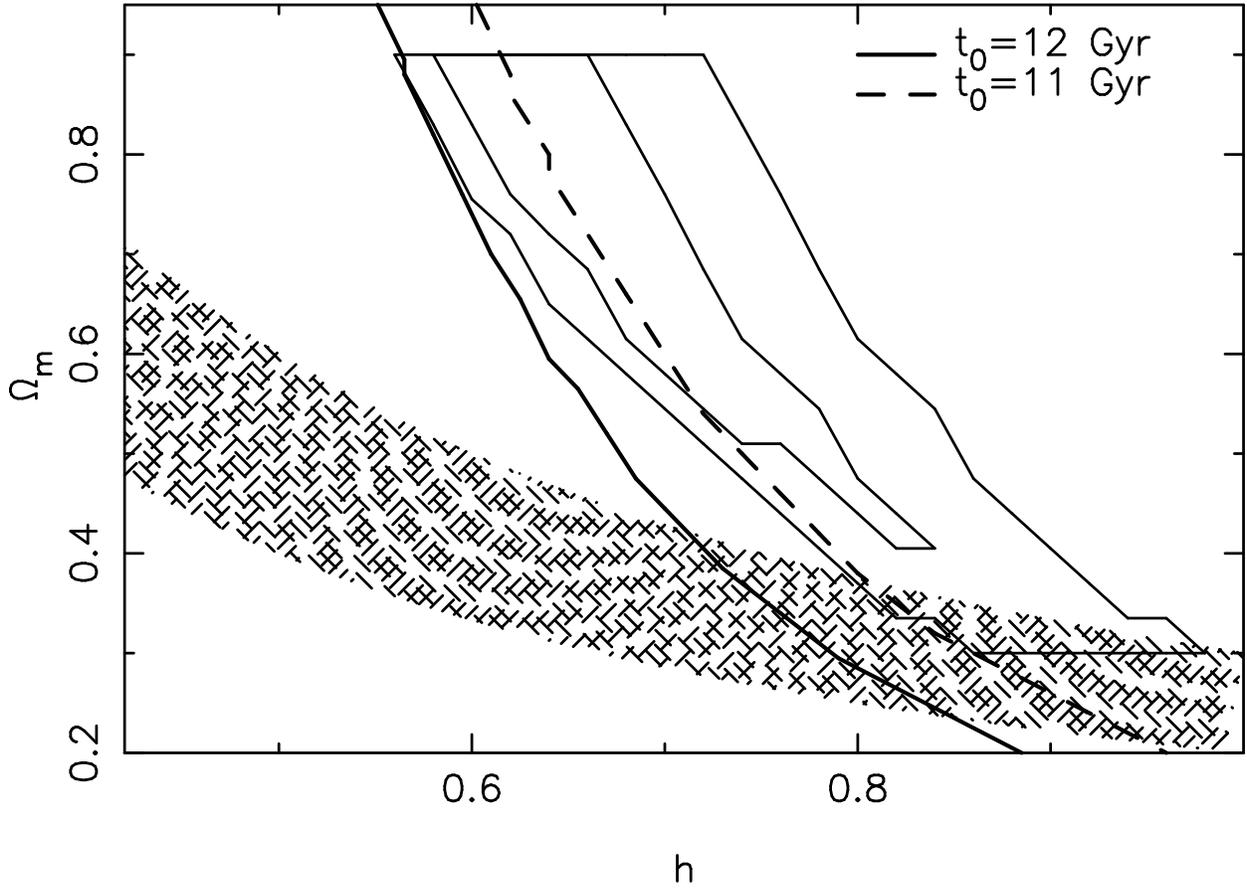}  
\caption{The contours correspond to allowed 1 and 2 $\sigma$
  regions by CMBR observations (see text for detail). The hatched
  region corresponds to $\Omega_m h = 0.25 \pm 0.05$. Other   curves show the
  age of the universe $t_0$ (value indicated in the figure legend)\label{fig:2}}
\end{figure}

\begin{figure}
\figurenum{3}
\includegraphics[angle=270, width=\textwidth]{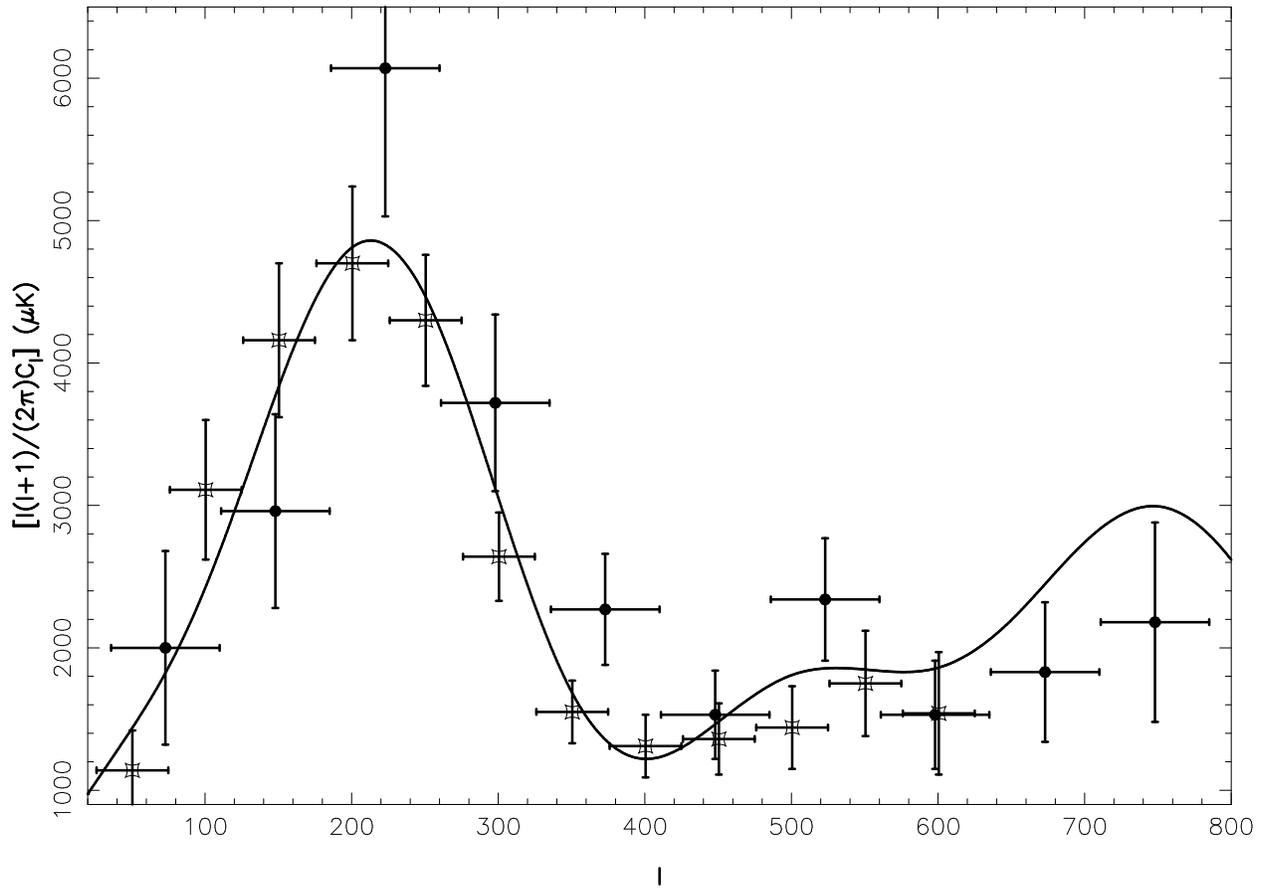}  
\caption{The best fit model is plotted along with BOOMERANG and MAXIMA data
  points, which correspond to empty and filled polygons, respectively.}
\end{figure}
 \end{document}